\documentclass[aps,superscriptaddress,showpacs,longbibliography,prl,twocolumn,floatfix]{revtex4-2}

\usepackage[utf8]{inputenc}
\usepackage{slashed}
\pdfoutput=1

\usepackage{color}
\usepackage{graphicx}   % Include figure filed
\usepackage{bm}
\usepackage{amsmath}
\usepackage{amsfonts}
\usepackage{hyperref}
\def\be{\begin{equation}}
\def\ee{\end{equation}}
\def\ba{\begin{eqnarray}}
\def\ea{\end{eqnarray}}

\begin{document}

\title{On self-dualities for scalar $\phi^4$ theory}

\author{Paul Romatschke}
\affiliation{Institut für Theoretische Physik, TU Wien, Wiedner Hauptstraße 8-10, 1040 Wien, Austria}

\begin{abstract}
Scalar field theory is studied by constructing interacting saddle point expansions in the symmetric and broken phase, respectively. Focusing on analytically tractable saddle expansions, it is found that broken and symmetric phases are related by sign flip of the quartic coupling. 
\end{abstract}

\maketitle

\section{Introduction}

Dualities are successful concepts in physics, having led to new and unexpected insights into problems ranging from quantum mechanics to quantum gravity \cite{deharo2025dualitiesphysics}. In this article, I focus on self-dualities in scalar field theory, some of which have been known for several decades \cite{Chang:1976ek,Magruder:1976px}. These dualities relate two different descriptions  of the same theory rather than different theories, hence the name ``self-duality''. For the present work, I consider a symmetric description where $\phi \leftrightarrow -\phi$ is a symmetry, and a ``broken phase'' description where it is not.

Considerable understanding of self-dualities for scalar field theory has been gained by high-order weak-coupling solutions in low dimensions, where it can be shown that perturbative expansions in the broken and symmetric phase are equivalent to all orders, thus proving equivalence up to non-perturbative corrections \cite{Serone:2018gjo,Serone:2019szm,Sberveglieri:2020eko}.

The aim of the present work is to focus on precisely these non-perturbative aspects of scalar field theory dualities, and how they relate to the established weak-coupling interpretation. Guided by insights in how large N field theory can be organized non-perturbatively \cite{Romatschke:2023ztk,Romatschke:2024yhx}, the approach taken here will be to identify saddle points for both the symmetric and broken phase description of scalar field theory. It is worth pointing out that these saddle points have an interpretation in terms of resumming Feynman diagrams, cf. Ref.~\cite{Romatschke:2019rjk}.

The focus will be on qualitative and analytic understanding of the big picture structure of the phase diagram of scalar field theory, rather than quantitatively precise results. In technical lingo, I will be employing R1 level resummations, while resolution schemes up to R4 have been implemented \cite{Romatschke:2019rjk,Romatschke:2021imm}. For this reason, I caution that some results found in this work do not agree with even leading-order weak-coupling perturbation theory (this shortcoming may be overcome by employing for instance R2-level resummations as in Ref.~\cite{Romatschke:2024yhx}).

Even though quantitatively unreliable, the results derived in the present work will be shown to be qualitatively consistent with known results obtained by other methods, such as lattice field theory and high-order perturbation theory in $d<4$ \cite{Schaich:2009jk,Serone:2018gjo,Serone:2019szm}. The case of $d=4$ is particularly interesting and warrants a special discussion.

After the discussion of results for scalar field theories derived in this work, and commenting on similarities and differences with respect to the literature, I will conclude to offer my own interpretation of self-dualities, while a compilation of calculational details and worked-through special cases is relegated to the appendix.

\section{Calculation}

I consider a scalar field $\phi$ with Euclidean action
\be
S=\int dx \left(\frac{1}{2}\partial_\mu \phi \partial_\mu \phi +\frac{1}{2}m_B^2\phi^2+ \lambda_B \phi^4 \right)\,,
\label{action}
\ee
where $\lambda_B,m_B$ are the bare parameters of the theory. Employing a Hubbard-Stratonovic transformation allows to rewrite the theory in terms of a \textbf{symmetric saddle expansion} in the auxiliary field. As shown in the appendix, the free energy density in d-dimensions $\Omega_{R1}^{(d)}$ in the R1 resummation becomes
\be
\label{OR1sym}
\Omega^{(d)}_{R1}(m_B)=-\frac{\Gamma\left(-\frac{d}{2}\right)}{2(4\pi)^{\frac{d}{2}}}(M^2)^{\frac{d}{2}}-\frac{(M^2-m_B^2)^2}{48\lambda_B}\,,
\ee
where the symmetric pole mass squared $M^2$ is specified by the saddle-point condition 
\be
\label{symsaddle}
M^2=m_B^2+12 \lambda_B \Delta(0)\,,\ \Delta(0)=\frac{\Gamma\left(1-\frac{d}{2}\right)}{(4\pi)^{\frac{d}{2}}}M^{d-2}%(M^2)^{\frac{d}{2}-1}\,.
\ee

By contrast, the description of the broken phase is organized by posing $\phi(x)=\phi_0+\xi(x)$ and expanding the action in powers of the fluctuation field $\xi(x)$. The calculation is organized in terms of a \textbf{broken phase saddle expansion} in terms of $\phi_0$, with the free energy density $\tilde \Omega_{R1}^{(d)}$ in the R1 level resummation found to be 
\be
\label{tildeOR1}
\tilde \Omega^{(d)}_{R1}=-\frac{\Gamma\left(-\frac{d}{2}\right)}{2(4\pi)^{\frac{d}{2}}}(\tilde M^2)^{\frac{d}{2}}+\frac{\tilde M^4}{96 \tilde \lambda_B}+\frac{\tilde M^2 \tilde m_B^2}{24\tilde \lambda_B}-\frac{\tilde m_B^4}{48\tilde \lambda_B}\,,
\ee
where following Ref.~\cite{Serone:2019szm} all quantities in the broken phase expansion have been denoted by tildes. (Note that the action is identical to the symmetric phase action when $\tilde m_B=m_B, \tilde \lambda_B=\lambda_B$.) The saddle point condition for the broken phase is stated in terms of the pole mass squared $\tilde M^2$ of the fluctuation field $\xi$,
\be
\label{bpsaddle}
\tilde M^2=-2\tilde m_B^2-24 \tilde \lambda_B \tilde \Delta(0)\,,\quad \phi_0^2=\frac{\tilde M^2}{8\tilde \lambda_B}\,.
\ee
Contrasting (\ref{bpsaddle}) with the corresponding (\ref{symsaddle}) from the symmetric saddle, one notices a small but important difference, namely the sign in front of the coupling. This difference is not due to an calculational error or a consequence of the approximation scheme, but is in fact generic.

\section{Chang duality in $d=2$}

Let me now evaluate and compare the results for the free energy of the theory in two dimensions, where the behavior of the theory is well studied, cf. Refs.~\cite{Schaich:2009jk,Serone:2018gjo,Serone:2019szm}. Using dimensional regularization with $d=2-2\varepsilon$ with $\varepsilon\rightarrow 0^+$ one has for the symmetric saddle expansion
\be
\Omega^{(2)}_{R1}=\frac{M^2}{8\pi}\left(\frac{1}{\varepsilon}+\frac{\pi m_B^2}{3\lambda_B}+\ln\frac{\bar\mu^2 e^1}{M^2}\right)-\frac{(M^4+m_B^4)}{48\lambda_B}\,,
\ee
where $\bar\mu$ is the $\overline{\rm MS}$ renormalization scale parameter. 
The divergence in the parentheses is canceled by the mass renormalization condition
\be
\label{m2dren}
\frac{m_B^2}{\lambda_B}=-\frac{3}{\pi \varepsilon}+\frac{m_R^2(\bar \mu)}{\lambda_B}\,,
\ee
which matches previous results \cite{Schaich:2009jk,Serone:2018gjo,Serone:2019szm}. The renormalization condition leads to a running mass parameter $m_R(\bar\mu)$, which is given by
\be
m_R^2(\bar\mu)=\frac{3\lambda_B}{\pi}\ln\frac{\Lambda_{\overline{\rm MS}}^2}{\bar\mu^2}\,,
\ee
where $\Lambda_{\overline{\rm MS}}$ is the $\overline{\rm MS}$ scale, again consistent with previous results. Inserting these results into the expression for the free energy density from the symmetric saddle expansion leads to
\be
\Omega^{(2)}_{R1}=\frac{M^2}{8\pi}\ln\frac{\Lambda_{\overline{\rm MS}}^2 e^1}{M^2}-\frac{(M^4+m_B^4)}{48\lambda_B}\,,
\ee
where the pole mass $M$ is given as the solution to the saddle-point condition
\be
\label{SP1}
\frac{M^2}{\lambda_B}=\frac{3}{\pi}\ln \frac{\Lambda_{\overline{\rm MS}}^2}{M^2}\,.
\ee
Note that the solution $M$ for the symmetric phase saddle expansion is given in terms of a Lambert W function, and is real-valued for all values of $\lambda_B>0$.

Now changing attention to the broken phase expansion in two dimensions, the corresponding results simplifies to
\be
\tilde \Omega^{(2)}_{R1}=\frac{\tilde M^2}{8\pi}\left(\frac{1}{\varepsilon}+\frac{\pi \tilde m_B^2}{3\tilde \lambda_B}+\ln\frac{\bar\mu^2 e^1}{\tilde M^2}\right)+\frac{\tilde M^4-2 \tilde m_B^4}{96\tilde \lambda_B}\,,
\ee
which can be renormalized exactly as in the symmetric phase using (\ref{m2dren}). One finds
\be
\tilde \Omega^{(2)}_{R1}=\frac{\tilde M^2}{8\pi}\ln\frac{ \Lambda_{\overline{\rm MS}}^2e^1}{\tilde M^2}+\frac{\tilde M^4-2 \tilde m_B^4}{96\tilde \lambda_B}\,,
\ee
where the pole mass $\tilde M$ is given as the solution of the saddle point condition (\ref{bpsaddle})
\be
\label{SP2}
\frac{\tilde M^2}{\tilde \lambda_B}=-\frac{6}{\pi}\ln \frac{\Lambda_{\overline{\rm MS}}^2}{\tilde M^2}\,.
\ee

It is important to note that (\ref{SP2}) for the pole mass in the broken phase corresponds to (\ref{SP1}) for the pole mass in the symmetric phase \textbf{at negative coupling}
\be
\label{relation}
\lambda_B=-2\tilde \lambda_B\,.
\ee

The fact that the renormalization condition used in the symmetric and broken phase expansion was identical implies that also the value of $\Lambda_{\overline{\rm MS}}^2$ in (\ref{SP1}) and (\ref{SP2}) is the same. Hence one can combine the two equations (\ref{SP1}, \ref{SP2}) into a single equation
\be
\pi\frac{\tilde M^2+2M^2}{6\lambda_B}+\ln \frac{M^2}{\tilde M^2}=0\,,
\ee
where I have set $\tilde \lambda_B=\lambda_B$ as outlined above. This equation implies a duality between the parameters of the symmetric phase and the broken phase in d=2, known as Chang duality \cite{Chang:1976ek,Serone:2019szm}.

Let me close the discussion of the d=2 theory by offering the following (lengthy) remark. There are two solutions to condition (\ref{SP2}) in terms of the Lambert W function,
\be
\label{twosol}
\tilde M^2=-\frac{6 \tilde \lambda_B}{\pi}W_{-1,0}\left(-\frac{\pi \Lambda_{\overline{\rm MS}}^2}{6\tilde \lambda_B}\right)
\ee
which are real-valued as long as
\be
\label{2dcond}
\frac{\tilde \lambda_B}{\Lambda_{\overline{\rm MS}}^2}\geq \frac{\pi e}{6}\,,
\ee
and form a complex-conjugate pair otherwise.  

Thus, unlike the symmetric phase expansion, the broken phase expansion only has real-valued pole masses for a restricted range of bare parameters. Since both symmetric phase and broken phase expansion are supposed to describe the same theory, the expectation is that the theory is in one of the two broken phases described by the two solutions (\ref{twosol}) if (\ref{2dcond}) is fulfilled, and otherwise in the symmetric phase \cite{Serone:2019szm}.

However, since we have access to the free energy of both expansions, we can set $\tilde m_B^2=m_B^2$ and $\tilde \lambda_B=\lambda_B$ and study the difference in free energy between solutions. A quick calculation reveals that the principal branch of (\ref{twosol}) is the lower free energy solution for the broken phase for all (\ref{2dcond}), hence I will disregard the second solution (\ref{twosol}) as corresponding to a thermodynamically disfavored meta-stable phase in the following. Comparing symmetric and broken phases one can use the finite difference 
\be
\Delta \Omega^{(2)}=\Omega^{(2)}_{R1}-\tilde \Omega^{(2)}_{R1}=\frac{M^2-\tilde M^2}{8\pi}+\frac{M^4}{48\lambda_B}+\frac{\tilde M^4}{96\lambda_B}\,,
\ee
to decide which phase has lower free energy. (Note that in the literature sometimes a finite ``counter-term'' is introduced in $\tilde \Omega^{(2)}$ in order to enforce $\Delta \Omega^{(2)}=0$ for all $\lambda_B$ \cite{Serone:2019szm}. I see no need for this ad-hoc procedure.)

Using the explicit results for $M,\tilde M$ as a function of $\lambda_B$, one finds that indeed the symmetric phase has lower free energy for small $\frac{\lambda_B}{\Lambda_{\overline{\rm MS}}^2}$, but that the free energy difference vanishes at the critical coupling
\be
\label{lambdac2d}
\lambda_c\equiv \frac{\lambda_B}{\Lambda_{\overline{\rm MS}}^2}\simeq 1.69371 > \frac{e\pi}{6}\,.
\ee

Starting from the small coupling regime, the thermodynamically preferred phase is therefore given by the symmetric saddle expansion until a coupling value $\lambda_c$ of (\ref{lambdac2d}), after which the thermodynamically preferred phase is given by the principal solution of the broken phase saddle condition (\ref{twosol}).
At the critical coupling value (\ref{lambdac2d}), the pole mass squared in the symmetric phase from (\ref{SP1}) fulfills
\be
\frac{\lambda_B}{M^2}=\frac{\pi}{3 W_0\left(\frac{\pi}{3\lambda_c}\right)}\simeq 2.5527\,,
\ee
which should be compared to numerical values of $\frac{\lambda_B}{M^2}\simeq 2.7\pm 0.025$ for instance from lattice field theory \cite{Schaich:2009jk}.

\vspace*{-0.4cm}
\section{Magruder duality in d=3}

In three dimensions, the absence of logarithmic singularities in the R1 resummation scheme make the comparison between symmetric and broken phase particularly simple. For the symmetric phase one finds
\be
\Omega_{R1}^{(3)}=-\frac{M^3}{12\pi}-\frac{(M^2-m_B^2)^2}{48\lambda_B}\,,
\ee
with $M=-\frac{3\lambda_B}{2\pi}+\sqrt{\frac{9\lambda_B^2}{4\pi^2}+m_B^2}$  and $\frac{\tilde \lambda_B^2}{m_B^2}\geq -\frac{4\pi^2}{9}$, 
%\ba
%\Omega_{R1}^{(3)}&=&-\frac{M^3}{12\pi}-\frac{(M^2-m_B^2)^2}{48\lambda_B}\,,\nonumber\\
%M&=&-\frac{3\lambda_B}{2\pi}\pm\sqrt{\frac{9\lambda_B^2}{4\pi^2}+m_B^2}\,,
%\ea
and for the broken phase 
\be
\tilde \Omega_{R1}^{(3)}=-\frac{\tilde M^3}{12\pi}+\frac{\tilde M^4}{96\lambda_B}+\frac{\tilde M^2 \tilde m_B^2}{24\lambda_B}-\frac{\tilde m_B^4}{48\lambda_B}\,,
\ee
with $\tilde M=\frac{3\tilde \lambda_B}{\pi}\pm\sqrt{\frac{9\tilde \lambda_B^2}{\pi^2}-2\tilde m_B^2}$. 
As was the case for $d=2$, one finds that broken and symmetric phase calculations are related by a sign-flip in the coupling (\ref{relation}). In the literature, dualities in d=3 that relate the broken and symmetric phase are known as Magruder duality \cite{Magruder:1976px}, cf. Ref.~\cite{Sberveglieri:2020eko} for a recent discussion.

The broken phase saddle point condition results in real-valued solutions for $\tilde M$ only for $\frac{\tilde \lambda_B}{\tilde m_B}\geq \sqrt{\frac{2\pi^2}{9}}\simeq 1.48$,  in analogy to the condition (\ref{2dcond}) found in d=2, and the lower free energy solution is given by $\tilde M=\frac{3\tilde \lambda_B}{\pi}+\sqrt{\frac{9\tilde \lambda_B^2}{\pi^2}-2\tilde m_B^2}$. Setting again $\tilde \lambda_B=\lambda_B, \tilde m_B=m_B$, one can directly study
$\Delta \Omega^{(3)}\equiv \Omega_{R1}^{(3)}-\tilde \Omega_{R1}^{(3)}$ to decide which phase has lower free energy. Decreasing $\frac{\lambda_B}{m_B^2}$ from infinity, the dominating phase is the broken phase until a critical value of $\lambda_c\equiv \frac{\lambda_B}{m_B}\simeq 1.55>\sqrt{\frac{2\pi^2}{9}}$, where $\Delta \Omega^{(3)}=0$. For lower values of $\frac{\lambda_B}{m_B^2}$, the symmetric phase is the thermodynamically preferred phase. The result for $\lambda_c$ can be compared to other numerical or high-order analytic methods which report a phase transition value of $\lambda_c\simeq 1.07$ \cite{Sberveglieri:2020eko}. As outlined in the introduction, the difference in numerical values for $\lambda_c$ is expected because the R1 resummation is a low-order resummation scheme not intended for quantitative accuracy.

\vspace*{-0.4cm}
\section{Self duality in d=4}

In dimensions $d=4-2\varepsilon$, one has
\be
\Omega^{(4)}_{R1}=-\frac{M^4}{64\pi^2}\left(\frac{1}{\varepsilon}+\ln\frac{\bar \mu^2e^{\frac{3}{2}}}{M^2}\right)-\frac{(M^2-m_B^2)^2}{48\lambda_B}\,,
\ee
with $M^2=m_B^2-\frac{3 M^2 \lambda_B}{4\pi^2}\left(\frac{1}{\varepsilon}+\ln\frac{\bar \mu^2e^{1}}{M^2}\right)$, whereas the broken phase free energy is given by
\be
\tilde \Omega^{(4)}_{R1}=-\frac{\tilde M^4}{64\pi^2}\left(\frac{1}{\varepsilon}+\ln\frac{\bar \mu^2e^{\frac{3}{2}}}{\tilde M^2}\right)+\frac{\tilde M^4}{96\tilde \lambda_B}+\frac{\tilde M^2 \tilde m_B^2}{24\tilde \lambda_B}-\frac{\tilde m_B^4}{48\tilde \lambda_B}\,,
\ee
with $\tilde M^2=-2\tilde m_B^2+\frac{3 \tilde M^2\tilde \lambda_B}{2\pi^2}\left(\frac{1}{\varepsilon}+\ln\frac{\bar \mu^2e^{1}}{\tilde M^2}\right)$, 
where $\bar\mu$ is again the ${\overline{\rm MS}}$ renormalization scale.

Requiring that the pole masses $M,\tilde M$ are finite leads to the following renormalization conditions:
\ba
\frac{1}{\lambda_B}=\frac{1}{\lambda_R(\bar\mu)}-\frac{3}{4\pi^2 \varepsilon}, \quad \frac{m_B^2}{\lambda_B}=\frac{m_R^2(\bar\mu)}{\lambda_R(\bar\mu)}\equiv m^2\,,\\
\frac{1}{\tilde \lambda_B}=\frac{1}{\tilde \lambda_R(\bar\mu)}+\frac{3}{2\pi^2 \varepsilon}, \quad \frac{\tilde m_B^2}{\tilde \lambda_B}=\frac{\tilde m_R^2(\bar\mu)}{\tilde \lambda_R(\bar\mu)}\equiv \tilde m^2 \,.
\label{reno}
\ea
In particular, denoting again the ${\overline{\rm MS}}$ parameter as $\Lambda_{\overline{\rm MS}}$, the running renormalized couplings become
\be
\lambda_R(\bar\mu)=\frac{4\pi^2}{3\ln \frac{\bar\mu^2}{\Lambda_{\overline{\rm MS}}^2}}\,,\quad
\tilde \lambda_R(\bar\mu)=\frac{2\pi^2}{3\ln \frac{\tilde \Lambda_{\overline{\rm MS}}^2}{\bar\mu^2}}\,,
\label{runco}
\ee
again is related to the renormalized coupling in the broken phase by the sign-flip relation (\ref{relation}). Using (\ref{runco}), the renormalized free energies become
\ba
\Omega^{(4)}_{R1}&=&-\frac{M^4}{64\pi^2}\ln\frac{\Lambda_{\overline{\rm MS}}^2e^{\frac{3}{2}}}{M^2}+\frac{M^2 m^2}{24}\,, m^2=\frac{3 M^2}{4\pi^2}\ln\frac{\Lambda_{\overline{\rm MS}}^2e^{1}}{M^2}\nonumber\\
\tilde \Omega^{(4)}_{R1}&=&-\frac{\tilde M^4}{64\pi^2}\ln\frac{\tilde \Lambda_{\overline{\rm MS}}^2e^{\frac{3}{2}}}{\tilde M^2}+\frac{\tilde M^2 \tilde m^2}{24}\,,
\tilde m^2=\frac{3 \tilde M^2}{4\pi^2}\ln\frac{\tilde \Lambda_{\overline{\rm MS}}^2e^{1}}{\tilde M^2}\,.\nonumber
\ea
(Note that in these expressions I have used $\lim_{\varepsilon\rightarrow 0} m_B,\tilde m_B\rightarrow 0$, which follows from (\ref{reno})). Results for $M^2$ are found to agree with Monte Carlo simulations on the lattice, see appendix.

One thus finds that the free energy of the broken phase and of the symmetric phase are in fact identical after renormalization: $\Omega^{(4)}_{R1}=\tilde \Omega^{(4)}_{R1}$ if $\tilde \Lambda_{\overline{\rm MS}}^2=\Lambda_{\overline{\rm MS}}^2$ The same is true for the pole masses: $M=\tilde M$. 

Unlike the Chang and Magruder dualities in $d=2,3$ the broken phase therefore is exactly self-dual to the symmetric phase \textbf{for the sign-flipped coupling relation} (\ref{relation}) in d=4. The ``opposite'' sign counter-term in the broken phase (\ref{reno}) has in fact been calculated a long time ago in the one-loop approximation \cite[Eq.~(3.12)]{Weinberg:1973am}, but the necessity for opposite-sign running coupling was not discussed (for details, verify how the running coupling was defined in Ref.~\cite{Weinberg:1973am}).

The present calculation therefore suggests that in $d=4$, scalar field theory with a negative coupling in the symmetric phase is dual to scalar field theory with positive coupling in the broken phase, and vice versa. This may be an important hint on how scalar field theory in d=4 can avoid triviality theorems \cite{Aizenman:2019yuo,Romatschke:2023sce}.

The saddle-point condition has two solutions
\be
M^2=-\frac{4 m^2 \pi^2}{3 W_{-1,0}\left(-\frac{4 m^2 \pi^2}{3 e^1 \Lambda_{\overline{\rm MS}}^2}\right)}\,,
\ee
which are real-valued for $m^2\leq \frac{3 \Lambda_{\overline{\rm MS}}^2}{4\pi^2}$, and where the principal branch solution has lower free energy.

I currently do not have full understanding on the correct treatment of the broken phase saddle solution, and more work in the future is needed. Nevertheless I want to entertain several options:

One option is to use the relation (\ref{relation}) but set $\tilde m_B^2=m_B^2$ such that $\tilde m^2=-\frac{m^2}{2}$. Identifying $\tilde \Lambda_{\overline{\rm MS}}=\Lambda_{\overline{\rm MS}}$ then gives
\be
\tilde M^2=\frac{2 m^2 \pi^2}{3 W\left(\frac{2 m^2 \pi^2}{3 e^1 \Lambda_{\overline{\rm MS}}^2}\right)}\,,
\ee
for the broken phase saddle, which is real-valued for all $m^2\geq -\frac{3 e^2\Lambda_{\overline{\rm MS}}}{2\pi^2}$, and results in $\tilde \Omega^{(4)}_{R1}<\Omega^{(4)}_{R1}$ for $m\geq 0$.

Another option is to impose $\tilde \lambda_B=\lambda_B$, or
\be
\tilde \lambda_B = -\frac{4\pi^2 \varepsilon}{3}+{\cal O}\left(\varepsilon^2\right),\quad \tilde m_B^2=\tilde m^2 \lambda_B={\cal O}(\varepsilon)\,.
\ee
In this case, the saddle point condition for the broken phase implies $\tilde M^2=0, \tilde \Omega_{R1}^{(4)}=0$ when $\varepsilon\rightarrow 0$. Setting $\tilde \Lambda_{\overline{\rm MS}}=\Lambda_{\overline{\rm MS}}$, one has $\tilde \Omega^{(4)}_{R1}=\Omega^{(4)}_{R1}$ at $m_{\rm crit}^2\simeq 0.0626\Lambda_{\overline{\rm MS}}^2 < \frac{3 \Lambda_{\overline{\rm MS}}^2}{4\pi^2}$ for the critical mass parameter where the transition takes place. 
Yet other options are possible, for instance setting
\be
\tilde \Lambda_{\overline{\rm MS}}^2=\frac{\bar\mu^3}{\Lambda_{\overline{\rm MS}}}\,,
\ee
with suitably chosen $\bar\mu$. This last option seems preferred when interpreting the high-temperature limit of the $d=4$ results in terms of the broken-phase $d=3$ theory.

\section{Summary and Conclusions}
\vspace*{-0.1cm}

In the present work, I have analyzed scalar field theory using two different interacting saddle expansions, namely a symmetric saddle and a saddle where the symmetry $\phi \leftrightarrow -\phi$ is broken, respectively. It is possible to relate the results for the two different expansions, leading to a set of dualities in different dimensions.

Working with an analytically tractable approximation, I showed that previous results in the literature for $d=2,3$ are qualitatively recovered. Specifically, my results correctly suggest a phase transition from a symmetric to a broken phase, even though the critical value of the coupling obtained in this work is numerically inaccurate. The case of d=1 (quantum mechanics) treated in the appendix is curious: the present method is qualitatively wrong (predicting a phase transition when there is none), but numerically accurate in that it almost matches the lowest-lying eigenvalue of the Hamiltonian for all values of $m_B^2$.

My interpretation of the two saddle point expansion is that they are descriptions of \textbf{different phases} of the same theory. For fixed bare parameters, these phases differ on a non-perturbative level, and a comparison of their respective free energies can be used to determine which phase is thermodynamically preferred. This is akin to the presence of stable and meta-stable phases in matter. Even though both descriptions are derived from the same Lagrangian, the descriptions are mutually exclusive (see also the discussion in the appendix).

A key result found in the present treatment was that the physical properties of the broken phase could be related to those of the symmetric phase with flipped-sign coupling (\ref{relation}), regardless of the field theory dimension d. This result implies that (at least within the R1 resummation scheme) negative coupling field theory has a physics interpretation in terms of a symmetry broken phase with positive coupling, and vice versa. This finding is of particular interest for the case of $d=4$, even though the correct interpretation and handling of the broken phase will benefit from future work. In this context, particularly the non-perturbative renormalization in the broken phase beyond the R1-level resummation will be of interest, cf.~\cite{Romatschke:2024yhx,Nath:2025vhg}, as will be applications to finite temperature \cite{Gould:2019qek,Romatschke:2024cld}.

To summarize, scalar field theory exhibits sign-flip coupling dualities between symmetric and broken phases on the level of R1-resummation level saddles. More work is needed to decide if this information is relevant or not.

\vspace*{-0.4cm}
  \section{Acknowledgments}

  I would like to thank the theory group members of INFN Florence for stimulating discussions and their hospitality and the authors of Ref.\cite{Lowdon:2024atn}
 for providing the numerical data in Fig.~\ref{fig:lat}.
\quad\\ \newpage

  \onecolumngrid
\begin{appendix}

  \section*{Appendix: Calculational Details}

In this appendix, I collect calculational details for results presented in the main text, as well as additional material.
  
A useful identity for calculations is the free energy density for a scalar field in $d$-dimensions, which is defined in terms of the Euclidean action (\ref{action}) with $\lambda_B=0$. One finds
\be
\label{ofree}
\Omega^{(d)}_{\rm free}(m_B)=\frac{1}{2{\rm vol}}{\rm tr}\ln[-\partial_\mu^2+m_B^2]=-\frac{\Gamma\left(-\frac{d}{2}\right)}{2(4\pi)^{\frac{d}{2}}}(m_B^2)^{\frac{d}{2}}\,,
\ee
where $\rm vol$ is the volume of space-time. To access the interacting theory, one can use resummed perturbation theory, for instance in the form of R-level resummation \cite{Romatschke:2019rjk}, which organizes the theory as expansions around non-perturbative (interacting!) saddle points.

The calculation will be performed for the same action, but expanding around two different class of saddles. First, I perform the expansion around a \textbf{symmetric phase saddle}, e.g. one where the symmetry $\phi\leftrightarrow \phi$ is manifest. Then, I perform the equivalent expansion around a \textbf{broken phase saddle}, where the symmetry is broken. The interpretation of the calculational results can be found in the main text of this article.

\subsection{Expansion around symmetric phase saddle}

  The expansion around the symmetric saddle proceeds by using the mathematical identity
  \be
  \label{HS}
e^{-\lambda_B \phi^4(x)}=\int \frac{d\zeta}{4\sqrt{\lambda_B \pi}} e^{-\frac{\zeta^2}{16\lambda_B}+\frac{i \zeta \phi^2(x)}{2}}\,, 
\ee
where $\zeta$ is an auxiliary field to rewrite the action as
\be
\label{symexp}
S=\int dx \left(\frac{1}{2}\partial_\mu \phi \partial_\mu \phi +\frac{1}{2}(m_B^2+i \zeta) \phi^2 +\frac{\zeta^2}{16\lambda_B}\right)\,.
\ee

Writing $\zeta(x)=\zeta_0+\xi(x)$, the zeroth order R-level resummation consists of neglecting fluctuations $\xi(x)$ in the evaluation of the free energy. (This corresponds to the leading large N behavior of an N-component scalar field theory\cite{Linde:1976qh}, see \cite{Romatschke:2023ztk} for a recent pedagogical review.) The fact that $\zeta_0$ is constant makes the R0-level free energy analytically accessible, and one finds
\be
\Omega^{(d)}_{R0}(m_B,i\zeta_0)=-\frac{\Gamma\left(-\frac{d}{2}\right)}{2(4\pi)^{\frac{d}{2}}}(m_B^2+i\zeta_0)^{\frac{d}{2}}+\frac{\zeta_0^2}{16\lambda_B}\,.
\ee
Here $\zeta_0$ is still understood to be integrated over as $\int d\zeta_0 e^{-{\rm vol}\times \Omega}$, which for large volume can be solved by a saddle-point evaluation. The corresponding saddle-point condition that fixes the value of $\zeta_0$ is
\be
0=\frac{d\Omega}{d(i\zeta_0)}\,.
\ee

A slightly better resummation that is still analytically tractable is the R1-level resummation, which adds and subtracts the term $\int dx \frac{1}{2}\nu^2 \phi^2$ to the action $S$. The value of $\nu^2$ is then determined by requiring that the two-point function $\Delta(x)\equiv \langle \phi(x)\phi(0)\rangle$ absorbs all non-trivial one-loop contributions through $\nu^2$, which leads to \cite{Romatschke:2019rjk}
\be
\nu^2=8\lambda_B \Delta(0)\,.
\label{nuval}
\ee
Note that in perturbative approaches, the R1-level resummation corresponds to the mean-field approximation. The only difference is conceptual: R1 defines a non-trivial \textbf{interacting} saddle for the theory, whereas mean-field resummation just absorbs a certain class of Feynman diagrams for observables in an expansion around the \textbf{non-interacting saddle}.

The free energy density expression in R1 becomes \cite{Romatschke:2019rjk}
\be
\Omega^{(d)}_{R1}(m_B,i\zeta_0)=-\frac{\Gamma\left(-\frac{d}{2}\right)}{2(4\pi)^{\frac{d}{2}}}(M^2)^{\frac{d}{2}}+\frac{\zeta_0^2}{16\lambda_B}-2\lambda_B \Delta^2(0)\,,
\ee
where I introduced the short-hand notation
\be
M^2=m_B^2+\nu^2+i \zeta_0\,,
\ee
for the pole-mass squared. With $\nu^2$ fixed through (\ref{nuval}), the value of $\zeta_0$ is once again fixed through the saddle-point condition, which becomes
\be
0=\frac{d\Omega^{(d)}_{R1}}{d(i\zeta_0)}=\frac{1}{2}\Delta(0)-\frac{i\zeta_0}{8\lambda_B}.
\ee
Note that ${\rm tr}\frac{1}{-\partial_\mu^2+M^2}={\rm vol}\times \Delta(0)$ and that any derivatives such as $\frac{d\nu^2}{d (i\zeta_0)}$ have canceled \cite{Romatschke:2019rjk}. As a consequence, the saddle point condition gives $i\zeta_0=4 \lambda_B \Delta(0)=\frac{\nu^2}{2}$ and one obtains for the free energy in the symmetric phase Eq.~(\ref{OR1sym}) in the main text.

I conclude by pointing out that for the symmetric saddle
\be
\label{s1p}
\langle \phi \rangle =0\,,
\ee
as expected.

\subsection{Expansion around broken phase saddle}

In contrast to the previous approach, let me now pose
\be
\phi(x)=\phi_0+\xi(x)\,,
\ee
and expand the action in powers of the fluctuation field $\xi(x)$. Note that this  differs qualitatively from the symmetric saddle expansion discussed above. Expanding the action $S$ in powers of $\xi$, one finds
\be
S=S_0+S_2+S_I\,,
\ee
with
\be
\label{bsexpansion}
S_0=\int dx \left[\frac{\tilde m_B^2}{2}\phi_0^2+\tilde \lambda_B \phi_0^4\right]\,,\quad
S_2=\int dx \left[\frac{1}{2}\partial_\mu \xi \partial_\mu \xi+\frac{\tilde m_B^2\xi^2}{2}+6\tilde \lambda_B \phi_0^2 \xi^2\right]\,,\quad
S_I=\int dx \left[4 \tilde \lambda_B \phi_0 \xi^3+\tilde \lambda_B \xi^4\right]\,.
\ee

A crude semi-classical approach is to neglect $S_2,S_I$ and determine the value of $\phi_0$ via the saddle point condition of the free energy, $0=\frac{d\tilde \Omega}{d\phi_0}$. However, one can do considerably better than this crude approximation, namely by performing the equivalent of the R1-level resummation of the preceding section for the broken-phase saddle expansion. To this end, add and subtract the term $\int dx \frac{1}{2} \tilde \nu^2 \xi^2$ to the action and determine $\tilde \nu^2$ by requiring that \cite{Romatschke:2019rjk}
\be
\langle \xi(x)\xi(0)\rangle=\langle\xi(x)\left(1-S_I\right)\xi(0)\rangle\,,
\ee
which leads to
\be
\label{tilnu}
\tilde \nu^2=12\tilde \lambda_B \tilde \Delta (0)\,,
\ee
where $\tilde \Delta(x)\equiv \langle \xi(x)\xi(0)\rangle$. The corresponding R1-level free energy expression for the broken saddle then becomes
\be
\tilde \Omega_{R1}^{(d)}(\tilde m_B,\phi_0)=\frac{\tilde m_B^2\phi_0^2}{2}+\tilde \lambda_B \phi_0^4-\frac{\Gamma\left(-\frac{d}{2}\right)}{2(4\pi)^{\frac{d}{2}}}(\tilde M^2)^{\frac{d}{2}}-3\tilde \lambda_B \tilde \Delta^2(0)\,,
\ee
where I introduced the short-hand notation
\be
\tilde M^2=\tilde m_B^2+12 \tilde \lambda_B \phi_0^2+\tilde \nu^2\,,
\ee
for the pole-mass squared of the fluctuation field.

With $\tilde \nu^2$ fixed through (\ref{tilnu}), the value of $\phi_0$ is once again fixed through the saddle-point condition, which becomes
\be
0=\frac{d\tilde \Omega^{(d)}_{R1}}{d\phi_0}=2\phi_0\left[\frac{\tilde m_B^2}{2}+2\tilde \lambda_B \phi_0^2+6\tilde \lambda_B\tilde \Delta(0)\right]\,,
\ee
Note that any derivatives such as $\frac{d\tilde \Delta(0)}{d\phi_0}$ have canceled. As a consequence, one finds
\be
\phi_0^2=-\frac{\tilde m_B^2+12\lambda_B\tilde \Delta(0)}{4\tilde \lambda_B}\,,
\ee
which leads to (\ref{bpsaddle}) stated in the main text. Evaluating the free energy density using these results one finds the result (\ref{tildeOR1}) stated in the main text. Let me conclude by pointing out that for the broken phase saddle, the partition function is a sum of the two contributions $\pm \phi_0$:
\be
\label{summi}
Z=\sum_{\pm \phi_0}e^{-{\rm vol}}\times \tilde \Omega_{R1}^{(d)}\,,
\ee
such that in particular
\be
\langle \phi \rangle =\phi_0-\phi_0=0\,,
\ee
exactly as in the symmetric phase (\ref{s1p}).

\subsection{Case d=1 aka Quantum Mechanics}

\begin{figure}
  \includegraphics[width=.7\linewidth]{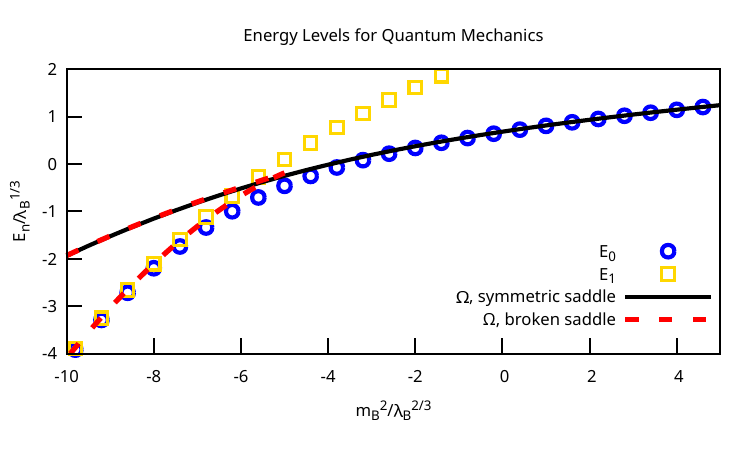}
  \caption{\label{fig:one} Comparison of lowest lying eigenvalues $E_0,E_1$ of the Hamiltonian to the free energies  $\Omega_{R1}^{(1)},\tilde \Omega_{R1}^{(1)}$ from the symmetric and broken saddles obtained in the R1-level resummation. See text for details.}
\end{figure}

An interesting test case is provided by considering d=1, which is the case of quantum mechanics. One has
\ba
\Omega_{R1}^{(1)}&=&\frac{M}{2}-\frac{(M^2-m_B^2)^2}{48\lambda_B}\,,\quad M^2=m_B^2+\frac{6 \lambda_B}{M}\,,\\
\tilde \Omega_{R1}^{(1)}&=&\frac{\tilde M}{2}+\frac{\tilde M^4}{96\lambda_B}+\frac{\tilde M^2 m^2_B}{24\lambda_B}-\frac{m_B^4}{48\lambda_B}\,,\quad \tilde M^2=-2m_B^2-\frac{12 \lambda_B}{\tilde M}\,,
\ea
where in the following I require the solutions to the saddle point equations $M,\tilde M \in \mathbb{R}^+$. (Note that this differs in philosophy from my complementary article \cite{Romatschke:2024cld}). The free energies $\Omega_{R1}^{(1)},\tilde \Omega_{R1}^{(1)}$ may then be directly compared to the eigenvalue spectrum for the quantum mechanical Hamiltonian ${\cal H}=-\frac{1}{2}\frac{d^2}{ dx^2}+V(x)$ with $V(x)=\frac{m_B^2x^2}{2}+\lambda_B x^4$, which can be obtained numerically. Discretizing the momentum operator using finite differencing on a space-lattice with $x_n=n a$, $n\in [-N,N]$, the Hamilton becomes matrix given by
\be
H=\left(\begin{array}{ccccc}
  \frac{1}{a^2}+V(x_{-N}) & -\frac{1}{2a^2} & 0 & 0 & \ldots\\
  -\frac{1}{2 a^2} & \frac{1}{a^2}+V(x_{-N+1}) & -\frac{1}{2a^2} & 0 & \ldots\\
  0 & -\frac{1}{2 a^2} & \frac{1}{a^2}+V(x_{-N+2}) & -\frac{1}{2a^2} & \ldots\\
  & & \ldots&& \\
  0 & 0 & \ldots & -\frac{1}{2a^2} & \frac{1}{a^2}+V(x_N)
\end{array}\right)\,,
\ee
  which is easily diagonalized numerically for given lattice spacing $a$ and sites $N$. The eigenvalues of $H$ correspond to the eigenvalues of the Hamiltonian in the limit $a\rightarrow 0$ and $a N\rightarrow \infty$. In practice, one chooses a given value for $a$ (say $a=0.1$) and then increases $N$ until the lowest-lying eigenvalues of $H$ have stabilized. Then $a$ is decreased and the process is repeated until the lowest-lying eigenvalues have become independent of the choice of $a, N$. Fig. \ref{fig:one} reports the lowest-lying eigenvalues $E_0,E_1$ after this numerical method has converged.

From Fig.~\ref{fig:one} one observes that the symmetric saddle solution for $\Omega_{R1}^{(1)}$ describes the ground state energy $E_0$ well for $m_B^2>0$, but is quantitatively inaccurate for $m_B^2<0$. For the broken phase, there are two solutions for $\tilde M\in \mathbb{R}^+$ if
\be
\frac{m_B^2}{\lambda_B^{\frac{2}{3}}}\leq -\left(\frac{243}{2}\right)^{\frac{1}{3}}\simeq -4.95\,.
\ee
The broken phase saddle $\tilde \Omega_{R1}^{(1)}$ for both of these solutions is shown in Fig.~\ref{fig:one}. As one can see, for one of these solutions $\tilde \Omega_{R1}^{(1)}$ approaches $\Omega_{R1}^{(1)}$ in the limit of $m_B^2\rightarrow -\infty$. This matches the qualitative behavior of the full theory, where $E_1$ and $E_0$ become degenerate in the limit of $m_B^2\rightarrow -\infty$, but there is no phase transition for finite $m_B^2$ (symbols in Fig.~\ref{fig:one}).  However, it can be seen from Fig.~\ref{fig:one} that the solution branches where the symmetric and broken phase free energies $\Omega_{R1}^{(1)}, \tilde \Omega_{R1}^{(1)}$ become degenerate is quantitatively different from the eigenvalues $E_0,E_1$ from numerical diagonalization.

Curiously, the behavior of $\tilde \Omega_{R1}^{(1)}$ for the \textit{second solution} $\tilde M$ (also shown in Fig.~\ref{fig:one}) matches the numerical eigenvalues $E_0,E_1$ quantitatively well for $m_B^2\rightarrow -\infty$. Since $\tilde \Omega_{R1}^{(1)}$ for this second solution is the lowest free energy solution, the present R1 saddle method wrongly predicts a phase transition from symmetric to broken phase at around $\frac{m_B^2}{\lambda_B^{\frac{2}{3}}}\simeq -5.4$, while offering a quantitatively accurate representation of the lowest lying eigenvalues of the Hamiltonian.

\subsection{Symmetric and Broken Saddle Expansions are Mutually Exclusive}

Starting from the broken saddle expansion (\ref{bsexpansion}), one might be tempted to rewrite the interaction $S_I$ using a mathematical identity similar to \ref{HS}. This procedure works, but completely undoes the expansion around a non-vanishing $\langle \phi\rangle$ as shown in the following. As a consequence, the symmetric saddle expansion and the broken phase saddle expansion are mutually exclusive and cannot be combined --- they correspond to different descriptions.

Inserting
\be
1=\int d\sigma \delta(\sigma-\xi^2)=\int d\sigma d\zeta e^{i\zeta (\sigma-\xi^2)}
\ee
into the path integral, the interaction part of (\ref{bsexpansion}) becomes
\be
S_I\rightarrow \int dx \left[4\tilde \lambda_B \phi_0 \xi \sigma+\tilde \lambda_B \sigma^2-i\zeta(\sigma-\xi^2)\right]\,,
\ee
which after integration over $\sigma$ gives for the action of the theory
\be
S=\int dx\left[\frac{m_B^2\phi_0^2}{2}+\tilde\lambda_B \phi_0^4+\frac{\xi}{2}\left(-\partial^2+m_B^2+4\tilde \lambda_B\phi_0^2+2 i \zeta\right)\xi+\frac{\zeta^2}{4\tilde \lambda_B}+2 i \phi_0 \zeta \xi\right]\,.
\ee
Shifting
\be
i\zeta\rightarrow i\zeta-2 \tilde \lambda_B \phi^2_0
\ee
leads to
\be
S=\int dx\left[\frac{m_B^2\phi_0^2}{2}+\frac{\xi}{2}\left(-\partial^2+m_B^2+2 i \zeta\right)\xi+\frac{\zeta^2}{4\tilde \lambda_B}+i\zeta \phi_0^2+2 i \phi_0 \zeta \xi\right]\,,
\ee
which can be rewritten as
\be
S=\int dx\left[\frac{(\xi+\phi_0)}{2}\left(-\partial^2+m_B^2+2 i \zeta\right)(\xi+\phi_0)+\frac{\zeta^2}{4\tilde \lambda_B}\right]\,.
\ee
Putting $\phi_0+\xi=\phi$, one recognizes the starting point for the symmetric saddle expansion (\ref{symexp}) in the main text. All dependencies on $\phi_0$ have disappeared. Therefore, the broken phase saddle expansion and symmetric saddle expansion cannot be combined.

\section{d=4 Comparison to Lattice Monte Carlo}

\begin{figure}[t]
  \includegraphics[width=.7\linewidth]{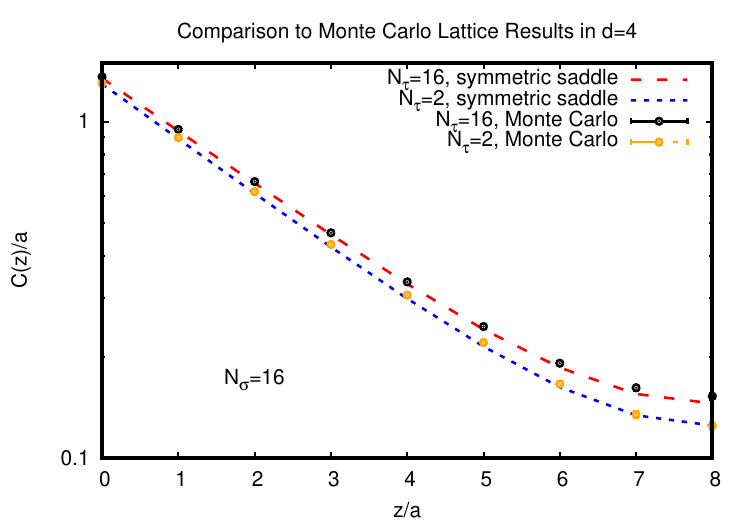}
  \caption{\label{fig:lat} Comparison of analytic saddle point results for the correlation function $C(z)$ to numerical lattice Monte Carlo results from Ref.~\cite{Lowdon:2024atn} for $m_B^2 a^2=0.15$ and $g_B=\frac{1.5}{4!}$. Shown are results for $N_\sigma=16$ and two different temporal lattice sizes $N_\tau=2,16$.  See text for details.}
\end{figure}

A non-trivial test of the performance of the saddle-point expansions is provided by comparison to Monte Carlo lattice data with positive bare coupling $\lambda_B>0$ in $d=4$. For the self-duality relations on the lattice, the main ingredient is the free energy for the free theory (\ref{ofree}) on a lattice with spacing $a$ and $N_\sigma,N_\tau$ number of points in the ``spatial'' and ``temporal'' Euclidean directions. One has
\be
\Omega_{\rm free}^{(4),{\rm lat.}}(m_B)=\frac{a^{-4}}{2 N_\sigma^3N_\tau}\sum_{n_1=1}^{N_\sigma}\sum_{n_2=1}^{N_\sigma}\sum_{n_3=1}^{N_\sigma}\sum_{n_4=1}^{N_\tau} \ln\left[\Omega_{N_\tau}^2(n_4)+\sum_{\mu=1}^{3}\Omega_{N_\sigma}^2(n_\mu)+m_B^2 a^2\right]\,,
\ee
where $\Omega_N^2(n)=2\left(1-\cos\left(\frac{2 \pi n}{N}\right)\right)$ are the lattice Matsubara frequencies squared. From this expression, one immediately obtains
\be
\Delta(0)=2 \frac{d}{dm_B^2}\Omega_{\rm free}^{(4),{\rm lat.}}(m_B)=\frac{a^{-2}}{N_\sigma^3N_\tau}\sum_{n_1=1}^{N_\sigma}\sum_{n_2=1}^{N_\sigma}\sum_{n_3=1}^{N_\sigma}\sum_{n_4=1}^{N_\tau} \frac{1}{\Omega_{N_\tau}^2(n_4)+\sum_{\mu=1}^{3}\Omega_{N_\sigma}^2(n_\mu)+m_B^2 a^2}\,,
\ee
so that the symmetric saddle saddle-point condition in R1 reads from (\ref{symsaddle})
\be
\label{latticegap}
M^2 a^2=m_B^2+\frac{12 \lambda_B}{N_\sigma^3N_\tau}\sum_{n_1=1}^{N_\sigma}\sum_{n_2=1}^{N_\sigma}\sum_{n_3=1}^{N_\sigma}\sum_{n_4=1}^{N_\tau} \frac{1}{\Omega_{N_\tau}^2(n_4)+\sum_{\mu=1}^{3}\Omega_{N_\sigma}^2(n_\mu)+m_B^2 a^2}\,.
\ee

For given bare lattice values $m_B^2,\lambda_B$, this equation may be directly solved to give the symmetric pole mass $M^2$ in lattice units for arbitrary lattice sizes $N_\sigma,N_\tau$.

Given a value of $M^2 a^2$, observables may be calculated that lend themselves to direct comparison to Monte Carlo simulations. One such observable that has been of interest recently is the scalar correlation function $\Delta(x)$ or more specifically on the lattice \cite{Lowdon:2024atn}
\be
C(z)=a^3 \sum_{x,y,\tau} \langle \phi(x)\phi(0)\rangle\,.
\ee

For the symmetric saddle in R1-resummation, this correlation function becomes
\be
C(z)=\frac{a}{N_\sigma}\sum_{n_3=1}^{N_\sigma} \frac{e^{\frac{2 \pi i n_3 z}{N_\sigma a}}}{\Omega_{N_\sigma}^2(n_3)+M^2 a^2}\,,
\ee
with $M^2 a^2$ given by the solution of (\ref{latticegap}). Results for $C(z)$ compared to the lattice data for the same quantity from Monte Carlo simulations from Ref.~\cite{Lowdon:2024atn} are shown in Fig.~\ref{fig:lat}. As can be seen from this figure, the analytic saddle point method is capable of quantitatively reproducing numerical lattice Monte Carlo results in four dimensions, without any need for fitting any parameters. This result is non-trivial since Ref.\cite{Lowdon:2024atn} provided evidence that weak-coupling perturbation theory fails qualitatively in describing the numerical Monte Carlo data, with the ordering of $C(z)$ for $N_\tau=2$ and $N_\tau=16$ reversed as compared to the Monte Carlo data.

\end{appendix}

\bibliography{enormous}
\end{document}